\documentclass[%
 twocolumn,
 jmp,%
 amsmath,amssymb,
 reprint,%
longbibliography]{revtex4-1}

\usepackage{xcolor}
\usepackage{float}
\usepackage{setspace}
\usepackage{subcaption}
\usepackage{graphicx}
\usepackage{dcolumn}
\usepackage{bm}
\newcommand{\RomanNumeralCaps}[1]
    {\MakeUppercase{\romannumeral #1}}
\makeatletter
\newcommand*{\rom}[1]{\expandafter\@slowromancap\romannumeral #1@}
\makeatother

\begin{document}

\title{{Horizontal component of the ion drag force in a plasma sheath}}


\author{S. Jaiswal}
\email{surabhi@iiserpune.ac.in}
\affiliation{Department of Physics, Indian Institute of Science Education and Research, Pune, India}
\author{S. A. Khrapak}
\affiliation{Joint Institute for High Temperatures, Russian Academy of Sciences, Moscow, Russia}
\author{E. M. Aguirre}
\affiliation{Plasma Unlimited LLC, Allenwood, Pennsylvania, USA}

\date{\today}

\begin{abstract}


\par The ion drag force is responsible for the underlying physics behind many important phenomena in complex (dusty) plasma, such as void formation, particle alignment in an ion flow, shadow-like attraction between the particles, dust clouds rotation in the presence of a magnetic field etc.  In this paper, we investigate the ion drag force acting on particles levitating near the sheath edge of a capacitively coupled discharge in xenon. In particular, we calculate the radial (horizontal) component of the ion drag force directed parallel to the lower electrode above which the particles levitate. The calculation is based on the actual information about the ion flow, which is directly obtained using Laser Induced Fluorescence. The calculated horizontal component is comparable in magnitude to the vertical one, which pushes the particles toward the electrode surface, and is higher than previously expected. This unexpected finding must be taken into account in future research. The very possibility of the measurements we report opens the door for detailed studies of dusty plasma phenomena governed by ions and their flows.  
\end{abstract}

\maketitle

\section{\label{sec:level1}Introduction}
\par The presence of charged macroparticles or ``dust", of nanometer to micrometer size, in a plasma environment is an ubiquitous occurrence in nature with examples including planetary rings, comet tails, interplanetary media, and interstellar clouds~\cite{Goertz}. Dusty plasmas are also important for numerous technological applications. For example, semiconductors are the backbone of modern technology, with plasma-based processes involved in nearly half of all semiconductor fabrication steps. In these processes, naturally formed dust particles are considered unwanted, and their control mechanisms are important to improve yield and device performance. Dusty plasmas in space are crucial for understanding satellite dust interactions~\cite{satellite_dust}, space based communication~\cite{dusty_plasma_radioblackout}, and space based manufacturing~\cite{dust_manufacturing}.
In dusty plasma systems, the characteristic length and time scales are significantly extended -- for example, to hundreds of microns and tens of milliseconds. This makes dusty plasmas an excellent medium for studying a wide range of generic processes operating in classical fluids and solids in exceptional detail, down to the most fundamental, “atomistic” (kinetic) level~
\cite{ThomasNature1996,KhrapakPRL2011,jaiswal_melting_DC, jaiswal_microgravity_ddw}. In recent years, this field has developed substantial interdisciplinary connections in areas such as statistical physics~\cite{statistical,KhrapakPRE09_2024}, soft condensed matter~\cite{condensed_matter,ChaudhuriSM2011}, physics of fluids~\cite{HuangPRR2023,KhrapakPhysRep2024}, and materials science.

\par Dust particles once introduced into the plasma acquire an electrical charge by collecting electrons and ions, interact with the plasma species, and self-consistently modify the surrounding plasma medium, making it more complex. Simultaneously, plasma species govern the behavior of dust particles and support a wide range of physical phenomena including self-organization \cite{TsytovichUFN1997,Kretschmer, melzer_crystal}, phase transitions in 2d and 3d complex plasma\cite{ThomasNature1996,KhrapakPRL2011,jaiswal_melting_DC}, waves~\cite{jaiswal_microgravity_ddw,SchwabeNJP2020,jaiswal_soliton,jaiswal_shock}, instabilities \cite{Rosenberg_ion_streaming,FortovPoP2000,MAMUN_SHUKLA_2011,FortovPoP2000}, void formation~\cite{GoreePRE1999,Kretschmer_void_microgravity,LipaevPRL2007}, sheared rotation, and much more. Ions are fundamental to all these phenomena. Therefore, accurate information of ion properties such as ion flow velocity and temperature will help develop the theoretical understanding of dust plasma interaction more accurately. 

\par 
The ion drag force is invoked for numerous dusty plasma phenomena, as explained above. It is also the least trivial force among the forces acting on charged particles in dusty plasmas and can vary greatly depending on plasma parameters and especially those of the ion component. Not surprisingly, it has been investigated extensively both theoretically and computationally. Most of the previous experimental investigations measured the ion drag force using indirect techniques. Some examples include measurements of the ion drag force from the particle mass and the angle of deflection as dust fell in the plasma medium~\cite{ion_drag_hirt1,zafiu2003,khrapak_comment_iondrag,nosenko_iondrag, beckers_iondrag}, or from the velocities of dust particles moving in a weak electric field of the positive column of a direct-current discharge~\cite{ion_drag_yaroshenko,WimmerNJP2025}. Very recently measurements of the ion drag force in the presheath region of single and dual-frequency capacitively coupled radio frequency discharge using optically trapped microparticles have been reported by Niemann \textit{et} al. ~\cite{NiemannPoP2025}. In this work the authors have calculated the localized ion drag force component by scanning the
particle vertically through the pre-sheath region. The horizontal motion was not measured. In general, experimental measurements of the ion drag force in the pre-sheath and sheath areas of a radio frequency discharge remain scarce.  


\par From the theoretical side, previous works on the ion drag force mainly utilized two approaches: the ``binary collision approach" and the ``kinetic approach". The former was discussed by Uglov and Gnedovets~\cite{uglov_binary_iondrag}; subsequent work was carried out by Barnes et al.~\cite{ion_drag_barnes} and Khrapak et al.~\cite{ion_drag_khrapak,khrapak_2003_ion_drag,KhrapakIEEE2004,KhrapakMT} taking into account specific properties of dusty plasma. In particular, the binary collision approach can account for strong nonlinearities in the ion-particle interaction, but it is unable to account for ion-neutral collisions and anisotropy in the ion-particle interaction induced by ion flows. For this reason, a self-consistent kinetic formalism was also developed~\cite{ion_drag_kinetic1, ion_drag_kinetic2,KhrapakJAP2007}. The kinetic approach is able to account for ion-neutral collisions and flow-induced anisotropies, but cannot account for strong nonlinearities is the ion-particle interaction since linear response technique is involved. Using this kinetic approach, the ion drag force was calculated for an arbitrary velocity of the ion flow~\cite{ion_drag_arbitrary}. A hybrid approach was developed simultaneously accounting for strong ion-particle coupling and anisotropy in the ion-particle interaction in a plasma with flowing collisionless ions~\cite{ion_drag_hybrid}. There is still no solid theoretical model for the regime in which ions are collisional and interact strongly with the particle (we do not consider this regime here). However, numerical simulations exist~\cite{PatacchiniPRL2008,HutchinsonPoP2013} and fitting formulas have been proposed.

\par In this paper, we utilize direct measurements of the intrinsic properties of the ion component, such as the drift velocity and temperature, to calculate the ion drag force acting on particles levitating near the sheath edge of a capacitively coupled discharge in xenon. Our ion velocity distribution function (IVDF) measurements via Laser Induced Fluorescence (LIF) provide the direct measurements of ion parameters that we use to calculate the component of the ion drag force acting in the radial direction (parallel to the lower electrode). Previously, most attention was directed to the vertical component, which pushes the particles to the electrode~\cite{ion_drag_barnes}, since it can affect the levitation height of the particles above the electrode. Our present results can help better understand the problems related to the formation and evolution of 2d dust voids in a radio-frequency plasma sheath~\cite{Dahiya2002} and, more generally, about the balance of different forces under sheath-related conditions. 

\section{Experimental Apparatus \& Methods}

\begin{figure}[htb]
    \centering
    \includegraphics[width=0.5\textwidth]{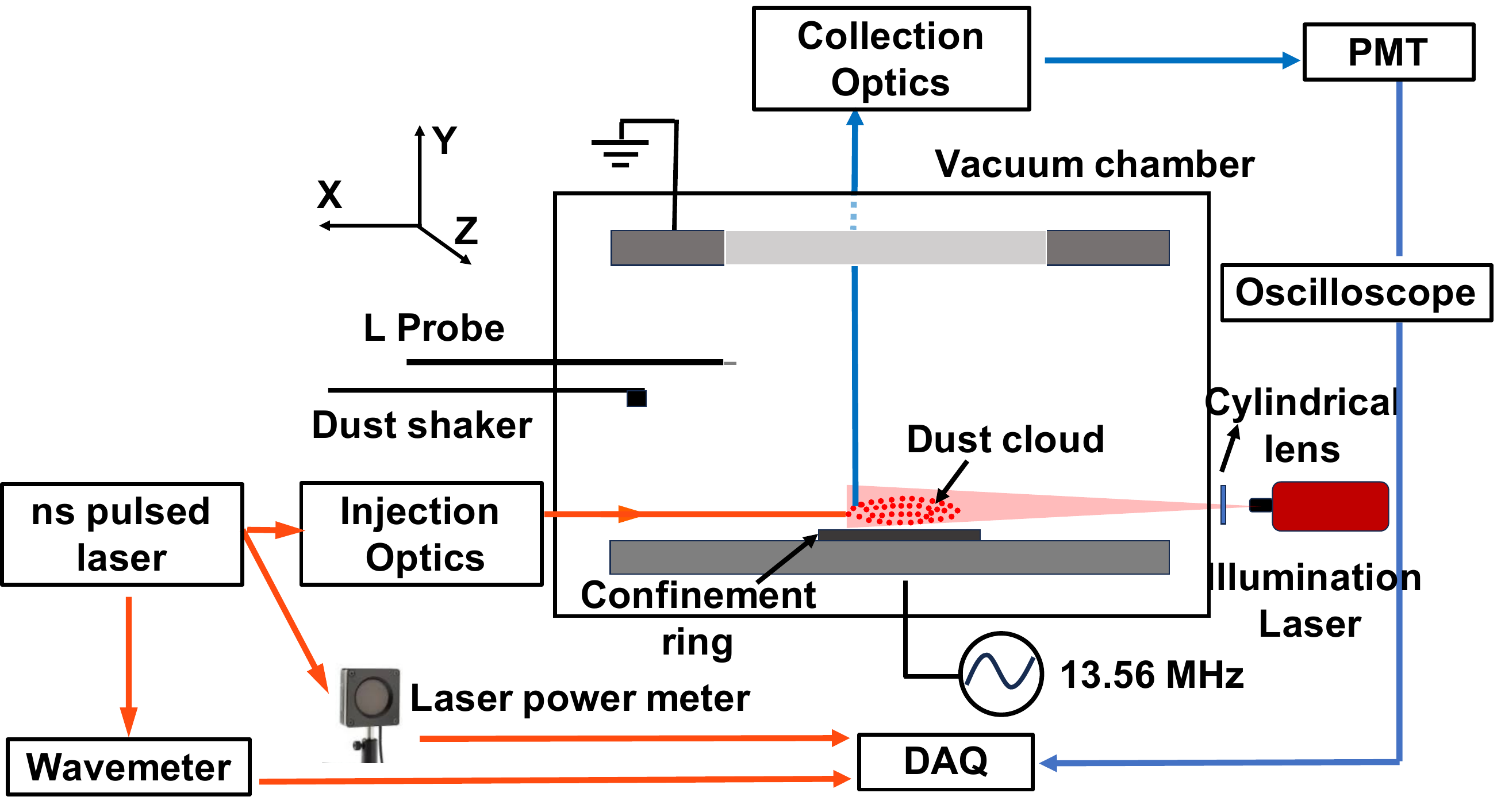}
    \caption{Schematic of the experimental arrangement. Levitated dust particles in xenon plasma over the rf powered electrode at 13.56 MHz are illuminated by a 650 nm laser. A ns-pulsed laser tuned at 605 nm was used for LIF measurements.}
    \label{fig:schematic}
\end{figure}

\indent A schematic of the experimental arrangement consisting of a dusty plasma source and the LIF system is shown in Fig.~\ref{fig:schematic}. A brief overview is given here as the detailed description of the LIF technique and plasma system is given elsewhere~\cite{Aguirre2026}. The experiment is performed in a  radio frequency (rf) capacitively coupled discharge (CCP) in a proprietary vacuum chamber and electrode system. An rf source at 13.56 MHz is coupled through a matching network to an electrode in the plasma chamber. The bottom electrode is powered by rf while the top electrode is grounded. The top electrode is made of conductive but transparent Indium Tin Oxide (ITO) which allows collection of light emission for LIF. Gas injection and vacuum pumping via a turbomolecular pump are arranged below the bottom electrode to create a stable and reproducible plasma. A Langmuir probe and dust shaker are attached via ports on one side of the chamber. The other sides of the chamber offer optical access for laser injection for LIF and laser illumination of the dust particles. A circular copper ring is used to trap the dust particles within a localized area. The dust particles are mono-disperse melamine-formaldehyde spheres of diameter 7.14$\pm 0.06~\mu$m.

\begin{figure}[htb]
    \centering
    \includegraphics[angle=0,width=0.5\textwidth]{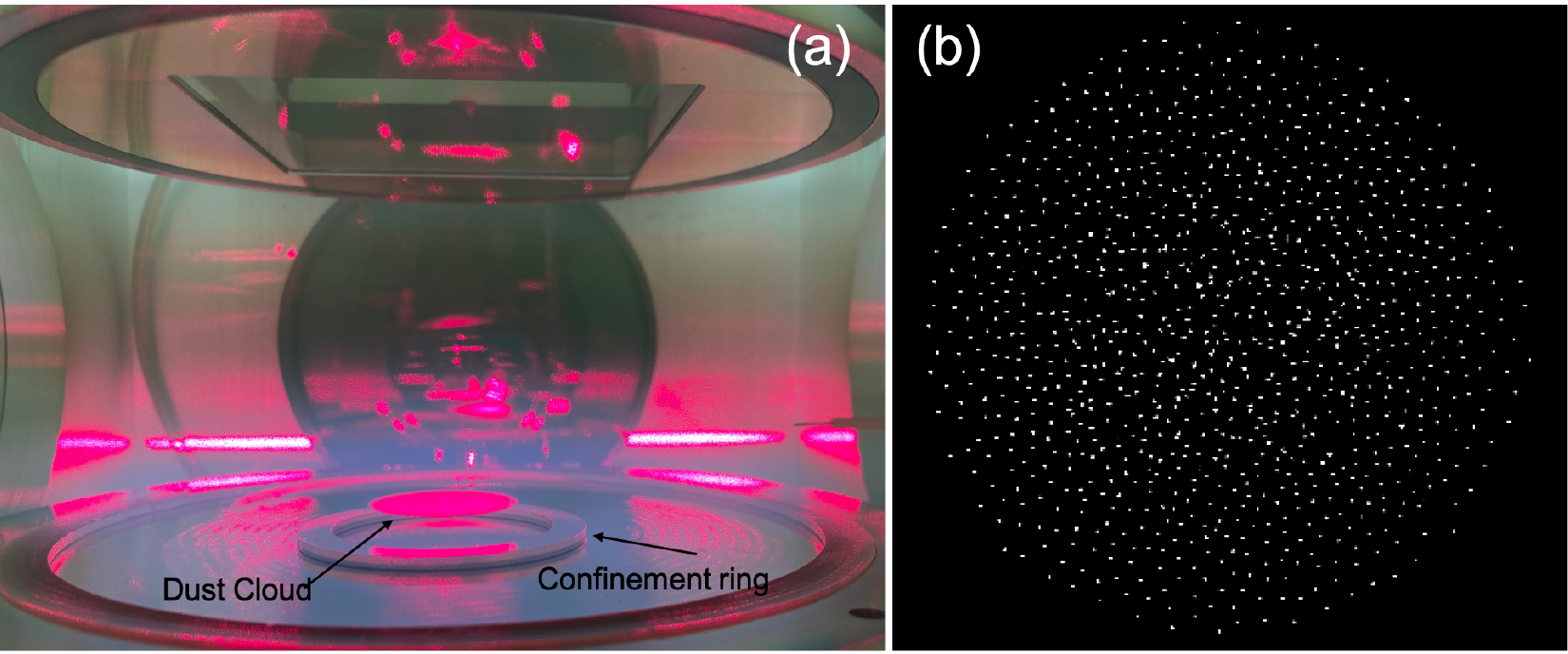}
    \caption{(a) Photograph of the dust cloud in the xenon plasma discharge for 10 W rf power and 8.25 mTorr pressure. (b) Top view of the dust cloud at these parameters.}
    \label{fig:dust_cloud}
\end{figure}

\indent A xenon plasma was created for a range of pressures and rf powers suitable for understanding dusty plasma phenomena. A photograph of the levitated dust particle cloud, illuminated with a continuous-wave laser at 650 nm is shown in Fig.~\ref{fig:dust_cloud}(a) with a monochrome top-view of the individual dust particles in Fig.~\ref{fig:dust_cloud}(b). Since LIF works well at low pressures, we investigated the lowest pressure where a stable dust cloud forms and worked into a higher pressure regime where dusty plasma phenomena are of interest. 

The LIF laser was injected and focused in the plane parallel to the electrode surface and is shown in the schematic of Fig.~\ref{fig:schematic}. The LIF measurement location is 2 cm from the center axis, near the edge of the dust cloud. All of our measurements are indicative of ions in this horizontal direction, also denoted by $v_{\perp}$, whereas the vertical direction of ions entering the plasma sheath is represented by $v_{\parallel}$. For the work presented here, a ns-pulsed tunable dye laser is tuned to 605.283 nm to pump the Xe \RomanNumeralCaps{2} 5$d^{4}D_{7/2}$ metastable state to the 6$p^{4}P_{5/2}$ state, which then decays to the 6$s^{4}P_{5/2}$ state by emitting 529.369 nm photons. A set of collection optics guides the overlapped excitation laser spot ($\sim 1$ mm) through a narrow bandpass filter centered on 529 nm placed before a photomultiplier tube (PMT) which is then connected to a 350 MHz bandwidth oscilloscope. A WS7-60 HighFinesse wavemeter, 60 MHz (0.08 pm) with $\pm~36$ m/s velocity error at 605 nm, monitors the laser wavelength. Data from the oscilloscope (i.e. PMT), wavemeter, and laser power meter, for measuring individual shot laser energy, are recorded by a custom data acquisition system.

\par The LIF spectral profile is subject to multiple line broadening mechanisms which must be taken into account to correctly measure the ion temperature. The ion flow is determined from the Doppler shift and is thus not altered by these mechanisms. Our deconvolution process is described in detail elsewhere~\cite{Aguirre2026} and we generally follow the procedure established by Smith \textit{et al.}~\cite{smith_thesis} to account for natural, hyperfine, Doppler, and laser linewidth contributions. 

\section{Results and Discussion}


\begin{figure}
    \centering
    \includegraphics[angle=0,width=0.45\textwidth]{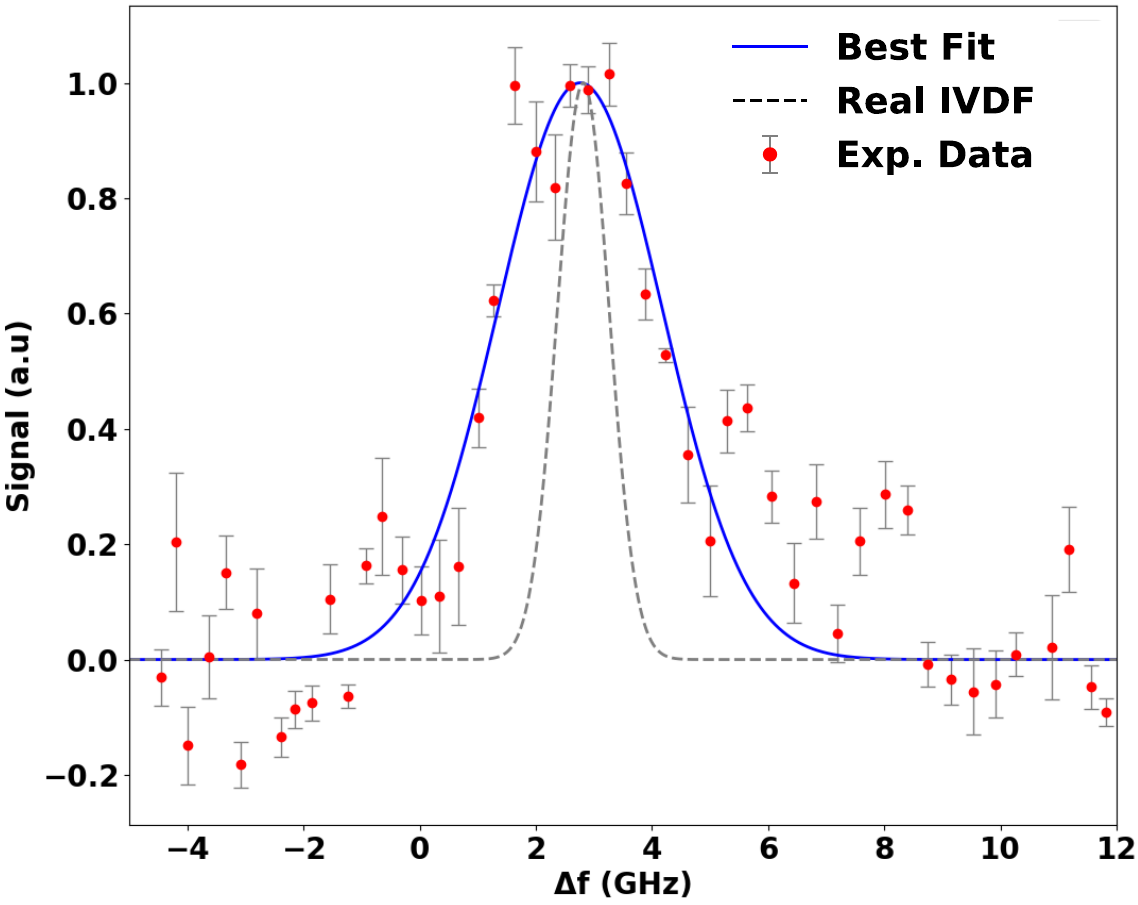}
    \caption{Ion velocity distribution function at the sheath interface for 10 W rf power and $p=8.25$ mTorr without dust present in the plasma discharge. The ion velocity is 1700 m/s and the ion temperature is 580 K.}
    \label{fig:ivdf_nodust}
\end{figure}

\indent
The ion velocity distribution function without dust at 8.25 mTorr and 10 W rf power is shown in Fig.~\ref{fig:ivdf_nodust}
The red data points are the measured IVDF, the solid blue line is the best fit for the simulated IVDF consisting of broadening mechanisms, and the dashed black line is the true IVDF from which the ion flow and ion temperature are extracted. For presentation purposes, the IVDF is normalized to the maximum signal. For a pressure of 8.25 mTorr and 10 W rf power, the real distribution of the ions yields clear values for $T_{i}\simeq580$ K and $v_{i}\simeq1700$ m/s. A Langmuir probe, in the absence of dust, measured the electron temperature and the electron density as  $T_{e}\simeq 3.1$ eV and $n_{e}\simeq 1.7 \times 10^{16}$ m$^{-3}$, respectively. The ion flow speed is slightly higher than the ion sound speed given by $v_{s}=\sqrt{T_{e}/m_{i}}\simeq1540$ m/s and much higher than the ion thermal speed given by $v_{Ti}=\sqrt{T_{i}/m_{i}}=190$ m/s (here $k_{\rm B}$ is set to unity to simplify the notation). 


\par The IVDF was also measured for the same plasma parameter as shown in Fig.~\ref{fig:ivdf_nodust} while dust was present. The addition of dust in the plasma reduces the ion flow velocity by $\sim$100 m/s, while the ion temperature remains the same within the error bar range. 

\subsection{The Ion Drag Force}

\par The parameter regime realized in this experiment provides a direct opportunity to calculate the ion drag force with input from our IVDFs. The situation is close to that considered by Barnes {\it at al.}~\cite{ion_drag_barnes}, except we look at the ion drag force acting in the horizontal direction, that is parallel to the electrode. 

For the low pressures investigated, the effect of ion-neutral collisions on either particle charging or ion drag force can be safely neglected. The fact that the ion drift velocity is comparable to the ion sound velocity has three important consequences. First, the ion-particle coupling is relatively weak. Second, the screening of the particle charge is mostly associated with electrons, since highly energetic ions barely contribute to the screening, see e.g. Ref.~\cite{ion_drag_hybrid} for a quantitative consideration. Third, since the drift velocity is much higher than the ion thermal velocity, the exact shape of the IVDF is not very important. The ion drift velocity can be decomposed into horizontal, $v_{\perp}$ which is the velocity component in the radial direction (shown as X-direction in Fig.~\ref{fig:schematic}), and vertical $v_{\parallel}$ components, which is the velocity component along the interelectrode gap and the discharge axis (shown as Y-direction in Fig.~\ref{fig:schematic}). The horizontal component is measured to be near-sonic $v_{\perp}\sim v_{\rm s}$. The vertical component should be comparable; the particles need to be levitated against gravity and therefore a significant electric field is required to ensure levitation. Such a strong electric field can only be produced near the sheath edge and, according to the Bohm criterion, $v_{\parallel}\sim v_{\rm s}$. Higher velocity can be realized deep in the sheath, but here the electron density vanishes exponentially, and sufficient negative charges on the particles cannot be sustained. The total drift velocity is therefore $v_{\rm tot}=\sqrt{v_{\parallel}^2+v_{\perp}^2}\sim \sqrt{2}v_{\rm s}$. To calculate the particle charge, we can apply the orbital-motion limited theory (OML)~\cite{Allen1992,FortovPhysRep2005}. We need to equate the electron and ion fluxes directed to the particle surface. The electron flux is
\begin{equation}
J_e = \sqrt{8\pi} a^2 n_ev_{Te}e^{-z},
\end{equation}
where $a$ is the radius of the particles, $n_e$ is the electron density, $v_{Te}=\sqrt{T_e/m_e}$ is the thermal velocity of the electrons, and $z=Ze^2/aT_e$ is the reduced particle charge, $Z$ being the charge number and $e$ the elementary (electron) charge.
The ion flux is 
\begin{equation}
J_i\simeq \sqrt{2}\pi a^2n_iv_{\rm s}(1+z),
\end{equation}
where $n_i$ is the density of the ions and we take into account $v_{\rm tot}\sim \sqrt{2}v_{\rm s}$. Equating the two fluxes, we get $z\simeq 4.6$. For our experiment, the radius of the particles is $3.57$ $\mu$m. The electron temperature of $3.1$ eV was measured with a Langmuir probe situated in the same plane as the LIF measurement and where the dust would levitate. This corresponds to the particle charge number $Z=z(aT_e/e^2)\simeq 3\times 10^4$ elementary charges. With the estimated particle density $n_p\simeq 1.5\times 10^5$ cm$^{-3}$ we obtain the charge density of the particle component $n_pZ\sim 4.5\times 10^9$. The electron density, also measured by the Langmuir probe, is 1.7$\times 10^{10}$ cm$^{-3}$. This implies that charge quasi-neutrality in the form $n_i\simeq n_e$ holds and the charge cannibalism effect~\cite{Khrapak_charge_cannibalism} is unimportant.

\par To calculate the absolute magnitude of the ion drag force, we multiply the Coulomb scattering cross section~\cite{ion_drag_khrapak,KhrapakPoP2025}
\begin{equation}
\sigma_{\rm C}=4\pi\left(\frac{Ze^2}{m_iv_{\rm tot}^2}\right)^2\ln\Lambda
\end{equation}
with the ion momentum flow $m_in_iv_{\rm tot}^2$. Here, $\ln\Lambda$ is the Coulomb logarithm. Generally, the Coulomb logarithm for ion-particle collisions in dusty plasma is a delicate topic and the standard approach either requires a modification or is not applicable at all due to strong ion-grain coupling~\cite{ion_drag_khrapak,khrapak_2003_ion_drag,KhrapakIEEE2004,KhrapakMT}. In this case, no such problem arises because the ions are highly energetic. To a good accuracy $\ln\Lambda\simeq \ln(\lambda_e/a)\sim 3$, where $\lambda_e$ is the electron Debye radius (screening is mostly associated with electrons in this case). We rewrite the ion drag force in the form
\begin{equation}\label{Fid}
F_i\simeq 2\pi z^2a^2n_eT_e\ln\Lambda,    
\end{equation}
where all the parameters are known. The force is composed of vertical $F_{\parallel}=F_i(v_{\parallel}/v_{\rm tot})$ and horizontal components $F_{\perp}=F_i(v_{\perp}/v_{\rm tot})$. In our case, these are comparable, $F_{\parallel}\sim F_{\perp}\sim 3\times 10^{-6}$ dyne. This is a noteworthy result since the horizontal component which was mostly neglected in past studies due to non availability of true ion velocity and temperature information is of same order of magnitude as the vertical component. This shows that different components of the ion drag force are important and should be taken into consideration in complex plasmas. The measured horizontal component of the ion drag force is also higher than in computational simulations of an rf discharge by Ikkurthi et al.~\cite{ikkurthi_iondrag}. The main reason is our higher density and lower operating pressure. Another interesting finding is that the magnitude of the total ion drag force is larger than the gravitational force. For example, the force of gravity acting on the particles is almost one order of magnitude smaller, $F_g\sim 3\times 10^{-7}$ dyne. This implies that particle levitation is dominated by the electric force acting upward and the ion drag force acting downward, rather than the force of gravity. From Eq.~(\ref{Fid}) it appears that this is mainly related to a relatively high energy density $n_eT_e$ of our plasma. 

\begin{table}[t]
  \centering
  \begin{tabular}{ | m{1.9cm} | m{1.5cm} | m{1.5cm} | m{1.2cm} | m{1.5cm} | }
 \hline
  $F_{i,\perp}$ (dyne) & Pressure (mTorr) & rf power (W) & $T_{i}$ (K) & $v_{i}$ (m/s)\\  
 \hline\hline
 3.3 $\times 10^{-6}$ & 6.15 & 10 & 550 & 1690\\ 
 \hline
 3.4 $\times 10^{-6}$ & 6.15 & 10 & 550 & 1740\\
 \hline
 2.3 $\times 10^{-6}$ & 8.25 & 10 & 500 & 1620\\
 \hline
\end{tabular}
\caption{Ion velocity and ion temperature from IVDFs and corresponding ion drag force values.}
\label{tab:ion_drag}
\end{table}
\par The horizontal component of the ion drag force over a range of a few mTorr and similar power levels is shown in Table~\ref{tab:ion_drag}. The ion drag force is influenced by the ion temperature and ion velocity. Because both the ion temperature and ion velocity are reduced at higher pressure, the ion drag force is also reduced slightly but the change is within the range of error bars. Therefore, over a range of a few mTorr and similar power levels, the horizontal ion drag force is essentially constant. Although there was a measurable change in the ion velocity when dust was not present, this change is not enough to affect the ion drag force considerably.

\section{Conclusion}
\indent In conclusion, we have presented the ion drag force calculation in xenon plasma in the presence and absence of dust particles. Although the ion drag force has previously been studied via various approaches
our direct measurement of ion parameters allowed us to calculate the ion drag force transparently from appropriate theory. Changes in ion flow and temperature affect the ion drag force by just a few percent, so it is usually a reasonable approximation to neglect the effect of dust on the ion flow. However, the order of magnitude of the ion drag force that we determined is considerably stronger than recent simulations of similar discharges~\cite{ikkurthi_iondrag}. Moreover, we have found that  
the horizontal component of the ion drag force is comparable to the vertical component which may have important consequences for a wide variety of phenomena such as, for instance, void formation in presheath and sheath plasma conditions. Moreover, we have demonstrated that there exist regimes in laboratory discharges where the force of gravity acting on micrometer-size particles is considerably smaller than the electric and ion drag force. Our result presented here will provide vital information to refine coupled ion dust modeling and pave the road for future experimental and theoretical studies in complex plasma.  




\section*{Acknowledgements}
The authors acknowledge support from the Center for KINETIC Plasma Physics at West Virginia University to conduct LIF measurement. E. M. A thanks Thomas Steinberger for his assistance in acquiring data and helpful discussions.

\section*{Data Availability}
The data that support the findings of this study are available from the corresponding author upon reasonable request.
%

\end{document}